\documentstyle[preprint,aps]{revtex}

\begin{document}
\tighten
\draft

%\twocolumn[\hsize\textwidth\columnwidth\hsize\csname
%@twocolumnfalse\endcsname

\title{Flat directions, doublet-triplet splitting, the monopole 
problem, and all that}

\author{Borut Bajc$^{(1)}$, Ilia Gogoladze$^{(2,3)}$, 
Ramon Guevara$^{(4)}$, and Goran Senjanovi\'c$^{(2)}$}

\address{$^{(1)}$ {\it J. Stefan Institute, 1001 Ljubljana, Slovenia}}
\address{$^{(2)}${\it International Center for Theoretical Physics,
Trieste, Italy }}
\address{$^{(3)}${\it Andronikashvili Institute of Physics, Georgian 
Academy of Sciences, 380077 Tbilisi, Georgia}}
\address{$^{(4)}${\it Dept. of Physics, University of Trieste, 34100 
Trieste, Italy}}

\maketitle

\begin{abstract}

We discuss a supersymmetric $SU(6)$ grand unified theory
with the GUT flat direction being lifted by soft supersymmetry
breaking, and the doublet-triplet splitting being achieved with 
Higgs as a pseudo-Goldstone boson. The theory offers a simple 
solution to the false vacuum and monopole problems.

\end{abstract}

\section{Introduction} 

The doublet-triplet splitting (D-T) problem and the origin of the 
unification scale are the outstanding problems of grand-unification. 
Both of them seem to cry for low energy supersymmetry which, 
miraculously enough, leads automatically to the unification of 
couplings
\cite{Dimopoulos:1981yj,Ibanez:1981yh,Einhorn:1982sx,Marciano:1982un}
and the dynamical generation of electroweak scale 
\cite{Alvarez-Gaume:1983gj}. 

Among various proposals to understand the lightness of the Higgs 
doublets, the mechanism that stands out 
is based on the beautiful idea of Higgs being pseudo-Goldstone 
boson of some accidental global symmetry
\cite{Inoue:1986cw,Anselm:1988um,Anselm:1988ss,Berezhiani:1989bd}. 
A particular simple realization of this scenario is realized in an
SU(6) GUT with an anomalous U(1)$_A$ symmetry \cite{Dvali:1997sr}. 

On the other hand the most elegant mechanism for generating the GUT 
scale seems to be based on the idea of flat directions
\cite{Witten:1981kv}, 
often naturally present in supersymmetric gauge theories. These flat 
directions are lifted after supersymmetry breaking and their large 
vevs can be traced to the logarithmic running of coupling constants 
and masses. 

In this paper we show how these appealing scenarios could be married 
in a realistic grand unified model. Our strategy is the following. 

As in \cite{Dvali:1997sr} we separate various sectors 
of the theory through U(1)$_A$ in order to maintain the lightness 
of the Higgs doublets. Next, in order to guarantee the existence of 
flat directions we employ (an) additional global symmetries(y). 
This is the easiest way to achieve the right pattern of symmetry 
breaking. 

On the other hand it is not so appealing to believe in 
global symmetries free from $1/M_{Pl}$ suppressed corrections. Thus 
we make an attempt to avoid completely global symmetries, i.e. to 
use only the local anomalous U(1)$_A$. While we cannot rigorously 
derive the correct symmetry breaking pattern in this case, we do 
believe that this is the most appealing possibility, worth 
pursuing further in future. If proven correct this would mean 
a realistic grand unified theory with a natural doublet-triplet 
splitting and the GUT scale determined dynamically. 

All this sounds nice. However, this program from its beginning 
suffered from a lack of phenomenological predictions and thus 
it becomes almost a question of semantics and not physics. 
Fortunately there is a possibility, as we show in this paper, 
that magnetic monopoles produced in the early universe are 
detectable in future experiments such as MACRO. The number of 
monopoles cannot be precisely calculated at this stage, but 
it could be comparable with the dark matter density of the 
universe. 

Last, but not least, we briefly comment on various other 
cosmological issues such as the false vacuum, gravitino and 
moduli problems. 

\section{A prototype model} 

Before presenting a realistic theory we wish to discuss the generic 
features of the lifting of flat directions. The simplest GUT example 
is based on SU(6) gauge symmetry with the adjoint representation $\Sigma$ 
and the following superpotential:

\begin{equation}
W={\lambda\over 3} Tr\Sigma^3\;.
\end{equation}

The absence of the mass term is simply a desire to determine masses 
dynamically and can be accounted by an appropriate R symmetry. It 
is clear that  the direction 

\begin{equation}
\Sigma=\sigma \; diag(1,1,1,-1,-1,-1)
\end{equation}

\noindent
is a flat direction since it disappears from the superpotential. 
It is also clear that this can only work in SU(2n) theories and 
thus not in SU(5). In this scenario one imagines the soft terms 
to originate at the Planck scale and to be positive as in the 
simplest models of supergravity. As in the MSSM the Higgs mass can 
change the sign \cite{Alvarez-Gaume:1983gj} and due to the larger 
number of fields this can now happen close to the GUT scale 
$M_{GUT}$ of the order $10^{16}$ GeV
\cite{Tabata:1983fz,Tabata:1984cr,Goldberg:1997ry,Dedes:1998px}. 

To complete the symmetry breaking down to the standard model 
the minimal set of Higgs scalars is a  
fundamental ($H$) and antifundamental ($\bar H$) representation. 
This can be achieved by nonrenormalizable terms in the superpotential 
or through D-terms. The latter case is preferred if one wants 
to avoid the introduction of arbitrary mass terms. An appealing 
possibility is to have $H=\bar H$ as a flat direction, but the 
trouble is the absence of enough running to change the sign of the 
soft mass terms at high enough scale. The way out is to introduce 
an extra (anomalous) gauge U(1)$_A$ symmetry, under which $H$ and 
${\bar H}$ are charged. A nonzero Fayet-Iliopoulos D-term 

\begin{equation}
\label{u1a}
D_{U(1)_A}=q_H|H|^2+q_{\bar H}|\bar H|^2+\xi=0
\end{equation}

\noindent
then forces nonvanishing (and equal) vevs for $H$ and $\bar H$:

\begin{equation}
\label{hhbar}
<H>=<\bar H>=\sqrt{-\xi\over q_H+q_{\bar H}}\;,
\end{equation}

\noindent
where from string theory

\begin{equation}
\xi={g^2 Tr Q\over 192\pi^2}M_{Pl}^2\;.
\end{equation}

What about the doublet-triplet splitting? Interestingly enough, it is 
achieved, but it ends up being a disaster: the SU(2) doublets are 
superheavy, while the colour triplets are light. Namely, if you do 
not couple $H$ and $\bar H$ to $\Sigma$, the F part of the 
potential has enlarged global symmetry SU(6)$_\Sigma\times$ 
SU(6)$_{H,\bar H}$. Let us imagine that $\Sigma$ first gets a 
vev, breaking the local SU(6)$\to$SU(3)$\times$SU(3)$\times$U(1). 
It is easy to see that all the particles of $\Sigma$ (except the 
flat direction $\sigma$) become superheavy. Now, let us trigger the 
vevs of $H$ and $\bar H$ so that we break one of the two remaining 
SU(3)'s down to SU(2). The doublet components of $H$ and $\bar H$ are 
obviously eaten by the corresponding gauge bosons, so that only 
the triplet components of SU(3) may (and do) remain light. This is 
confirmed by the counting of Goldstone bosons: 

\noindent
(i) SU(6)$_\Sigma\to$SU(3)$\times$SU(3)$\times$U(1): 
$35-(8+8+1)=18$;

\noindent
(ii) SU(6)$_{H,\bar H}\to$SU(5): $35-24=11$;

\noindent
(iii) SU(6)$\to$SU(3)$\times$SU(2)$\times$U(1): 
$35-(8+3+1)=23$;

\noindent
$18+11-23=6=3+\bar 3$.

The above example shows that it seems to be easier to find flat 
directions than to achieve natural doublet-triplet splitting. 
Therefore we now focus our attention on the model of D-T splitting 
which works and look for the implementation of flat directions. 

\section{A realistic theory}

What we learned in the previous example is that is not good to 
break SU(3) down to SU(2) with $H$ and $\bar H$, since the doublets 
get eaten and the SU(3) triplets remain light. We need SU(3) triplets 
to be eaten, and this can happen naturally when SU(4) is broken down 
to SU(3). In fact this is what Dvali and Pokorski do: they break 
SU(6) down to SU(4)$\times$SU(2)$\times$U(1) through the vev of 
$\Sigma$. At the next stage $H$ and $\bar H$ break SU(4), which as 
we said, makes the SU(3) triplets eaten and allows for the doublets to 
be light. A simple counting of Goldstone bosons demonstrates that the 
doublets are really light. 

Of course, the order of symmetry breaking is irrelevant for the above
arguments; if anything in supersymmetry one prefers to go through the
$SU(5)$ stage, i.e. to have first $H$ and $\bar H$ develop vevs (or
simultaneously with $\Sigma$).

It is easy to achieve the desired symmetry breaking \cite{Dvali:1997sr}; 
it is enough to choose the complete superpotential for $\Sigma$:

\begin{equation}
W={\lambda\over 3}Tr\;\Sigma^3+{m\over 2}Tr\;\Sigma^2\;.
\end{equation}

\noindent
One of the degenerate vacua is then 

\begin{equation}
\Sigma={m\over\lambda}\;diag(1,1,1,1,-2,-2)\;.
\end{equation}

The question of course is how to make it flat. The simplest 
possibility is to promote $m$ into a dynamical variable, i.e. 
a singlet field $S$. The trouble is that $F_S=0$ will make 
$\sigma$ vanish. Of course one can add a cubic self-interaction 
for $S$, but the equations $F_{\Sigma}=F_S=0$ over determine the 
system, forcing again the vevs to vanish. Notice that we are not 
allowed to introduce quadratic terms with our philosophy of 
generating masses dynamically. 

We see then that unfortunately the prize for achieving both the 
flatness and D-T splitting is to double the number of adjoints. 
Regarding the flat directions the situation here mimics the one 
encountered in SU(5) \cite{Goldberg:1997ry}. It is the D-T 
splitting problem that points to the elegant solution which 
requires SU(6) symmetry. 

\subsection{The model}

A simple model that implements our program requires two 
adjoint superfields $A$, $B$ and two singlet ones $S$, $S'$ 
with the following renormalizable superpotential 

\begin{equation}
W=\lambda_ATrA^2B+\lambda_SSTrAB+\lambda_{S'}S'TrB^2\;.
\end{equation}

A physical minimum of the potential is given by 

\begin{equation}
<A>={\lambda_S\over\lambda_A}<S>diag(1,1,1,1,-2,-2)\;\;,\;\;<B>=0\;\;,
\end{equation}

\noindent
with $<S>$ and $<S'>$ undetermined. 

The global symmetries of this superpotential are a U(1) R-symmetry 
and a U(1) global symmetry with charges $(1,1,1,1)$ and 
$(1,-2,1,4)$, respectively, for $(A,B,S,S')$, which forbids 
all other terms to all orders in $1/M_{Pl}$. 

One is clearly tempted to get rid of one of the singlet 
superfields, for example $S'$. This is readily achieved with 
$\lambda_{S'}=0$. This is a disaster for gauge coupling 
unification since both (4,2) and ($\bar 4$,2) multiplets 
under SU(4)$\times$SU(2) subgroup of SU(6) from $B$ would 
remain light. Of course you could add a term such as 
$TrAB^2$, but then no symmetry could forbid the 
$TrA^3$ term, which spoils flatness. 

Strictly speaking one can do without a U(1) R-symmetry. The reason 
is that only $B$ field carries a negative U(1) global charge and 
thus the nonrenormalizable terms will involve at least two powers 
of $B$. The vev of $B$, as is readily seen, still remains zero and 
the flatness is not spoiled. 

The Higgs as a pseudogoldstone boson program requires, as we 
mentioned before and as is well known, a separation in the 
superpotential of various sectors of the theory (for a systematic 
and careful study of this issue as a perturbation in powers of 
$1/M_{Pl}$ see for example \cite{Berezhiani:1995sb}). 
In particular $H$ and $\bar H$ must decouple 
from $A$, $B$, $S$ and $S'$. This can be 
achieved simply by giving nonzero and not opposite U(1)$_A$ 
charges only to $H$ and $\bar H$ as in \cite{Dvali:1997sr}. 

\subsection{Fixing the scales}

A few words are in order regarding the determination of 
$<A>$ which defines the GUT scale. Since the couplings 
$\lambda_{A,S,S'}$ are not known, the GUT scale cannot be 
determined from the first principles. However, since the 
number of fields in $A$ and $B$ is large compared to the 
situation in the MSSM, it is not surprising, that the running 
from $M_{Pl}$ down may be speeded up enough in order to flip 
the sign of the soft mass of the flat direction around the GUT scale. 
Furthermore, $A$ is also coupled to matter fields \cite{Barbieri:1994kw} 
and this can only help. For more details on similar models see 
\cite{Tabata:1983fz,Tabata:1984cr,Goldberg:1997ry,Dedes:1998px}. 

In summary, it appears, at least in the high energy sector of the 
theory, that everything works. The important ingredient though was 
at least one continuous global symmetry. Strictly speaking this 
is OK since we do not know the fate of global symmetries in 
the presence of quantum gravity. It is often suspected that 
only gauge symmetries are protected from gravitatonal, i.e. 
$1/M_{Pl}$-like effects. If one took this seriously it would be 
impossible to speak of Peccei-Quinn symmetry and the axion 
solution to the strong CP problem 
\cite{Kamionkowski:1992mf,Holman:1992us,Barr:1992qq}. 

\subsection{No global symmetries?}

Still, it would be reassuring to be able to get us rid of 
continuous global symmetries. It would also be much more 
elegant and physical to do so. The simplest and most appealing 
possibility is to use only the gauge (anomalous) U(1)$_A$. 
Actually, this could work in principle. Namely, in this case 
the U(1)$_A$ charges of ($A$, $B$, $S$, $S'$) would be 
(1,-2,1,4) instead of zero, and the $H$, $\bar H$ charges 
should be large enough and positive. As before, the fact that 
only $B$ has negative charge guarantees that the mixing between 
the two sectors involves more powers of $B$. For example, 
if the charges of ($H$, $\bar H$) are (2,2), the lowest 
order mixing would be proportional to $B^2H\bar H$, which 
is not harmful, since the vev of $B$ vanishes. 

There is however a new potential problem. In the original version, 
since only $H$ and $\bar H$ has nonvanishing U(1)$_A$ charges 
and since the SU(6) D-term has to be zero before supersymmetry 
breaking, both of these fields are forced to have nonvanishing 
and equal vevs (see (\ref{u1a}-(\ref{hhbar})). Now on the other 
hand the U(1)$_A$ D-term takes the form 

\begin{equation}
D_{U(1)_A}=Tr(AA^\dagger)-2Tr(BB^\dagger)+|S|^2+4|S'|^2+
q_H|H|^2+q_{\bar H}|\bar H|^2+\xi=0
\end{equation}

\noindent
and the issue who and when gets a vev becomes somewhat tricky. In 
order to answer this question the RG improved effective potential 
should be calculated using the running from $M_{Pl}$ to $M_{GUT}$. 
This is a difficult task beyond the scope of this 
paper. 

\subsection{The matter sector}

The theory can be made realistic with the proper inclusion of light 
matter superfields. A realistic theory can be shown to require three 
families of $15_f$, $\bar 6_f$ and $\bar 6'_f$ (a minimal anomaly 
free set). Also, one needs a self-conjugate $20$ of SU(6) in order 
to get a large top Yukawa coupling. This is discussed at length in 
\cite{Barbieri:1994kw,Shafi:2000tn}. A particular attention must 
be paid to neutrino mass as in general SU(6) models. Fortunately 
one has more than one option at disposal. Right-handed neutrinos 
can be the SU(5) singlet components of the $\bar 6$ and $\bar 6'$ 
matter fields as for example in \cite{Dvali:1997sr} or additional 
SU(6) singlets as in \cite{Shafi:2000tn}. In both cases one ends 
up with the usual mechanism for generation of small neutrino 
masses \cite{Mohapatra:1981yp}. 

\section{Cosmological issues: the monopole problem and 
the problem of the false vacuum}

Besides the well known monopole problem, SUSY GUTs are also 
plagued by the problem of the false vacuum. Namely, normally 
one gets a set of degenerate vacua which includes the unbroken 
one. At sufficiently high temperature the unbroken vacuum 
becomes the global minimum and the large barrier between the 
vacua prevents the tunneling to our world \cite{Weinberg:1982id}. 

The theories with flat directions offer a natural solution to 
both of these problems. First, the monopole problem. The point is 
remarkably simple
\cite{Klinkhamer:1982ga,Ginsparg:1982tq,Pi:1982fs,Yamamoto:1983ck}: the
critical temperature of the GUT 
phase transition becomes very small and the usual Kibble
\cite{Kibble:1976sj} 
mechanism production gets suppressed. On top of that, the phase 
transition is of the first order and the number of monopoles can get 
suppressed. For a small flat direction $\sigma$ ($<<T$) the 
one-loop high temperature correction to the effective potential is 

\begin{equation}
\label{dvt0}
\Delta V_T\approx -NT^4+\alpha T^2|\sigma|^2\;,
\end{equation}

\noindent
where $N$ is proportional to the degrees of freedom to which $\sigma$ 
is coupled and $\alpha$ is positive. In the opposite limit, when 
$\sigma>>T$, $\Delta V_T\approx exp(-c|\sigma|/T)$ ($c>0$), i.e. in this 
limit $\sigma$ is coupled only to superheavy fields ($>>T$) and is 
out of thermal equilibrium. Thus, for sufficiently high $T$ the $\sigma
= 0$ minimum wins and the symmetry is restored just as in the case
with no flat directions \cite{Haber:1982nb,Mangano:1984dq,Bajc:1996kj}.

Since the energy difference between the  
$\sigma=0$ and $\sigma=M_{GUT}$ vacua is only of order  
$m_{3/2}^2M_{GUT}^2$, it is clear that the transition can take place 
not before the temperature drops down to at least 
$T_c\approx (m_{3/2}M_{GUT})^{1/2} \approx 10^{9}-10^{10}$ GeV.  
If the phase transition was of the second order, the ratio between 
the energy of monopoles and baryons today would be approximately 

\begin{equation}
\left({\rho_M\over\rho_B}\right)_{today}\approx{m_M (n_M/n_\gamma)\over
m_B (n_B/n_\gamma)}\approx \left({T_c\over M_{Pl}}\right)^3\times 
10^{10}\times {m_M\over m_B}\approx 10^{-3} - 1\;,
\end{equation}

\noindent
for the GUT monopoles with a mass of the order $10^{17}$ GeV. 
Clearly, even if this was true, the number of monopoles would
be small enough not to be in conflict with cosmology. At first
glance, the usual curse of grandunification would be turned 
into the blessing: monopoles could be the dark matter of the
universe. Even more important, this is not far from the MACRO 
limit \cite{Adams:1993fj} and the old dream of detecting 
magnetic monopoles could be relized in not so far future. 

In our case the phase transition is of first order and the 
monopole production could be suppressed, although not 
necessarily (see for example 
\cite{Guth:1980bh,Guth:1981uk,Turner:1992tz}).

Of course, all this is relevant if we do manage to tunnel into our 
world. In a sense, we are saying that the solution to the false 
vacuum problem automatically resolves the monopole problem. 
The quasi flat direction may imply no barrier at all and so no 
problem whatsoever. However, this is in principle model dependent. 
Also, it is conceivable that the production of monopoles happens 
only after the false vacuum stops being a local minimum, i.e. 
for $T\approx m_{3/2}$. Obviously, the number of monopoles 
could then be completely negligible, similar to the case of 
inflation \footnote{We thank Gia Dvali for emphasizing this point.}.
A more careful study of these issues is on its way. 

\section{Summary and outlook}

In short, the $SU(6) \times U(1)_A$ theory discussed here achieves the
determination of the GUT scale through the lifting of the flat
direction after supersymmetry breaking. Also, it allows for a simple
solution of the doublet-triplet splitting problem with the Higgs
being a pseudo-Goldstone boson of an accidental global symmetry.

Furthermore, independently of when the inflation takes place and 
what the reheating temperature is, the theory is free from the 
monopole and false vacuum problems. Of course, we believe that 
inflation did take place at some point for the usual reasons of 
horizon and flatness problems. By this we mean the usual 
inflationary scenario of at least 60 e-folds, but now with a 
reheating temperature higher than $M_{GUT}$. 

For this reason one must face the gravitino problem. This is 
however easily solved by assuming a short inflation before 
the first order GUT phase transition discussed throughout this 
paper. It is enough to wash out the gravitinos thermally produced 
before it: the reaheating temperature will be smaller or at least 
equal to the critical temperature $T_C\approx 10^9$ GeV, which 
is safe. 

The main physical implication of flat directions is the 
existence of moduli-like fields with masses of order 
$m_{3/2}$ and $1/M_{Pl}$ suppressed interactions. It is well 
known (for the original work see \cite{Coughlan:1983ci}) that 
it poses a serious cosmological problem. It can be solved 
with a short inflation (this time after the phase transition) 
as suggested in \cite{Randall:1995fr,Dvali:1995mj,Dine:1995uk}. 

To be honest, both problems could be more severe through the 
non-thermal production of relics, as emphasized 
\cite{Kallosh:2000jj,Giudice:1999yt}. If so, one would need a 
low scale inflation at a later stage. The issue however is very 
subtle and recently an opposite point of view was raised 
\cite{Nilles:2001fg}, according to which the non-thermal 
production is suppressed in realistic models. 

The reader may feel uneasy about this multi-inflation scenario. 
We do not believe one should worry about it, since inflation 
is a natural scenario and often it is more of a problem to 
get out of it \cite{Guth:1983pn} than to experience it. 
In our scenario there is an award of having a possibility 
of detecting magnetic monopoles. This is the major point of 
our work. Magnetic monopoles as much as 
proton decay, if not more, provide a test of the idea of 
grand unification. After all, these are the only generic 
properties of GUTs. Of course, the usual inflation with low 
reheating temperature solves the monopole problem, but at the 
tragic prize of implying no monopoles left in the whole universe. 

\acknowledgments

We are grateful to Lotfi Boubekeur for collaboration in the early 
stage of this work and to Gia Dvali for important comments regarding 
the monopole production in theories with first order phase transition. 
The work of B.B. is supported by the Ministry of Education, Science 
and Sport of the Republic of Slovenia; the work of I.G. and G.S. is 
partially supported  by EEC under the TMR contracts ERBFMRX-CT960090 
and HPRN-CT-2000-00152. Both B.B. and R.G. thank ICTP for hospitality 
during the course of this work.

\end{document}